%% file: main.tex
\pdfoutput=1

\documentclass[12pt,a4paper]{article}

\usepackage{ifthen} 
\newboolean{pdflatex}
\setboolean{pdflatex}{true} 

\newboolean{articletitles}
\setboolean{articletitles}{true} 

\newboolean{uprightparticles}
\setboolean{uprightparticles}{false} 

\newboolean{inbibliography}
\setboolean{inbibliography}{false} 

\input{preamble}
\usepackage{longtable} 

\begin{document}

\renewcommand{\thefootnote}{\fnsymbol{footnote}}
\setcounter{footnote}{1}

\input{title-LHCb-PAPER}


\renewcommand{\thefootnote}{\arabic{footnote}}
\setcounter{footnote}{0}



\pagestyle{plain} 
\setcounter{page}{1}
\pagenumbering{arabic}


%

\input{introduction}

\input{detector}

\input{selection}

\input{massfit}

\input{systematics}

\input{bdlimit}

\input{summary}

\input{acknowledgements}

\addcontentsline{toc}{section}{References}
\setboolean{inbibliography}{true}
\bibliographystyle{LHCb}
\bibliography{main,LHCb-PAPER,LHCb-CONF,LHCb-DP,LHCb-TDR,local}

\newpage

\newpage

\input{LHCb_authorlist.tex}

\end{document}

%% file: preamble.tex
\usepackage{microtype}
\usepackage{lineno}  
\usepackage{xspace} 
\usepackage{caption} 

\usepackage{graphicx}  
\usepackage{color}
\usepackage{colortbl}
\graphicspath{{./figs/}} 

\usepackage{amsmath} 
\usepackage{amssymb}
\usepackage{amsfonts}
\usepackage{upgreek} 

\newcommand*\patchAmsMathEnvironmentForLineno[1]{%
\expandafter\let\csname old#1\expandafter\endcsname\csname #1\endcsname
\expandafter\let\csname oldend#1\expandafter\endcsname\csname
end#1\endcsname
 \renewenvironment{#1}%
   {\linenomath\csname old#1\endcsname}%
   {\csname oldend#1\endcsname\endlinenomath}%
}
\newcommand*\patchBothAmsMathEnvironmentsForLineno[1]{%
  \patchAmsMathEnvironmentForLineno{#1}%
  \patchAmsMathEnvironmentForLineno{#1*}%
}
\AtBeginDocument{%
\patchBothAmsMathEnvironmentsForLineno{equation}%
\patchBothAmsMathEnvironmentsForLineno{align}%
\patchBothAmsMathEnvironmentsForLineno{flalign}%
\patchBothAmsMathEnvironmentsForLineno{alignat}%
\patchBothAmsMathEnvironmentsForLineno{gather}%
\patchBothAmsMathEnvironmentsForLineno{multline}%
\patchBothAmsMathEnvironmentsForLineno{eqnarray}%
}

\usepackage{hyperref}    
\usepackage[all]{hypcap} 

\input{lhcb-symbols-def} 
\input{commands}

\usepackage{cite} 
\usepackage{mciteplus}

%% file: lhcb-symbols-def.tex
\def\lhcb {\mbox{LHCb}\xspace}

\def\MagUp {\mbox{\em Mag\kern -0.05em Up}\xspace}

\ifthenelse{\boolean{uprightparticles}}%
{

 \def\PDelta      {\ensuremath{\Delta}\xspace}                 
 \def\PXi      {\ensuremath{\Xi}\xspace}                 
 \def\PLambda      {\ensuremath{\Lambda}\xspace}                 
 \def\PSigma      {\ensuremath{\Sigma}\xspace}                 
 \def\POmega      {\ensuremath{\Omega}\xspace}                 
 \def\PUpsilon      {\ensuremath{\Upsilon}\xspace}                 
 
 \def\PB      {\ensuremath{\mathrm{B}}\xspace}                 
                  
 \def\PD      {\ensuremath{\mathrm{D}}\xspace}

 \def\PK      {\ensuremath{\mathrm{K}}\xspace}

 \def\Pb      {\ensuremath{\mathrm{b}}\xspace}                 
 \def\Pc      {\ensuremath{\mathrm{c}}\xspace}

 \def\Pi      {\ensuremath{\mathrm{i}}\xspace}

 \def\Pp      {\ensuremath{\mathrm{p}}\xspace}

 \def\Ps      {\ensuremath{\mathrm{s}}\xspace}

}
{

 \mathchardef\PDelta="7101
 \mathchardef\PXi="7104
 \mathchardef\PLambda="7103
 \mathchardef\PSigma="7106
 \mathchardef\POmega="710A
 \mathchardef\PUpsilon="7107
                  
 \def\PB      {\ensuremath{B}\xspace}                 
                  
 \def\PD      {\ensuremath{D}\xspace}

 \def\PK      {\ensuremath{K}\xspace}

 \def\Pb      {\ensuremath{b}\xspace}                 
 \def\Pc      {\ensuremath{c}\xspace}

 \def\Pi      {\ensuremath{i}\xspace}

 \def\Pp      {\ensuremath{p}\xspace}

 \def\Ps      {\ensuremath{s}\xspace}

}

\makeatletter
\ifcase \@ptsize \relax
  \newcommand{\miniscule}{\@setfontsize\miniscule{4}{5}}
\or
  \newcommand{\miniscule}{\@setfontsize\miniscule{5}{6}}
\or
  \newcommand{\miniscule}{\@setfontsize\miniscule{5}{6}}
\fi
\makeatother

\DeclareRobustCommand{\optbar}[1]{\shortstack{{\miniscule (\rule[.5ex]{1.25em}{.18mm})}
  \\ [-.7ex] $#1$}}

\def\squark    {{\ensuremath{\Ps}}\xspace}

\def\cquark    {{\ensuremath{\Pc}}\xspace}

\def\bquark    {{\ensuremath{\Pb}}\xspace}

\def\kaon    {{\ensuremath{\PK}}\xspace}
  \def\Kbar    {{\kern 0.2em\overline{\kern -0.2em \PK}{}}\xspace}

\def\KorKbar    {\kern 0.18em\optbar{\kern -0.18em K}{}\xspace}

\def\Kstar   {{\ensuremath{\kaon^*}}\xspace}
\def\Kstarb  {{\ensuremath{\Kbar{}^*}}\xspace}

  \def\Dbar    {{\kern 0.2em\overline{\kern -0.2em \PD}{}}\xspace}
\def\D       {{\ensuremath{\PD}}\xspace}

\def\DorDbar    {\kern 0.18em\optbar{\kern -0.18em D}{}\xspace}

\def\Dstar   {{\ensuremath{\D^*}}\xspace}

\def\B       {{\ensuremath{\PB}}\xspace}
\def\Bbar    {{\ensuremath{\kern 0.18em\overline{\kern -0.18em \PB}{}}}\xspace}

\def\BorBbar    {\kern 0.18em\optbar{\kern -0.18em B}{}\xspace}
\def\Bz      {{\ensuremath{\B^0}}\xspace}

\def\Bd      {{\ensuremath{\B^0}}\xspace}
\def\Bs      {{\ensuremath{\B^0_\squark}}\xspace}
\def\Bsb     {{\ensuremath{\Bbar{}^0_\squark}}\xspace}

  \def\Y#1S{\ensuremath{\PUpsilon{(#1S)}}\xspace}

\def\proton      {{\ensuremath{\Pp}}\xspace}

\def\Lz          {{\ensuremath{\PLambda}}\xspace}
\def\Lbar        {{\ensuremath{\kern 0.1em\overline{\kern -0.1em\PLambda}}}\xspace}
\def\LorLbar    {\kern 0.18em\optbar{\kern -0.18em \PLambda}{}\xspace}

\def\Lb      {{\ensuremath{\Lz^0_\bquark}}\xspace}

\def\to                 {\ensuremath{\rightarrow}\xspace}

\def\CP                {{\ensuremath{C\!P}}\xspace}

\def\AT#1     {\ensuremath{A_{\mathrm{T}}^{#1}}\xspace}           

\def\C#1      {\ensuremath{\mathcal{C}_{#1}}\xspace}                       
\def\Cp#1     {\ensuremath{\mathcal{C}_{#1}^{'}}\xspace}                    
\def\Ceff#1   {\ensuremath{\mathcal{C}_{#1}^{\mathrm{(eff)}}}\xspace}        
\def\Cpeff#1  {\ensuremath{\mathcal{C}_{#1}^{'\mathrm{(eff)}}}\xspace}       
\def\Ope#1    {\ensuremath{\mathcal{O}_{#1}}\xspace}                       
\def\Opep#1   {\ensuremath{\mathcal{O}_{#1}^{'}}\xspace}                    

\newcommand{\tev}{\ifthenelse{\boolean{inbibliography}}{\ensuremath{~T\kern -0.05em eV}\xspace}{\ensuremath{\mathrm{\,Te\kern -0.1em V}}}\xspace}
\newcommand{\gev}{\ensuremath{\mathrm{\,Ge\kern -0.1em V}}\xspace}
\newcommand{\mev}{\ensuremath{\mathrm{\,Me\kern -0.1em V}}\xspace}
\newcommand{\kev}{\ensuremath{\mathrm{\,ke\kern -0.1em V}}\xspace}
\newcommand{\ev}{\ensuremath{\mathrm{\,e\kern -0.1em V}}\xspace}
\newcommand{\gevc}{\ensuremath{{\mathrm{\,Ge\kern -0.1em V\!/}c}}\xspace}
\newcommand{\mevc}{\ensuremath{{\mathrm{\,Me\kern -0.1em V\!/}c}}\xspace}
\newcommand{\gevcc}{\ensuremath{{\mathrm{\,Ge\kern -0.1em V\!/}c^2}}\xspace}
\newcommand{\gevgevcccc}{\ensuremath{{\mathrm{\,Ge\kern -0.1em V^2\!/}c^4}}\xspace}
\newcommand{\mevcc}{\ensuremath{{\mathrm{\,Me\kern -0.1em V\!/}c^2}}\xspace}

\def\mum  {\ensuremath{{\,\upmu\rm m}}\xspace}

\def\invfb   {\ensuremath{\mbox{\,fb}^{-1}}\xspace}

\newcommand{\stat}{\ensuremath{\mathrm{\,(stat)}}\xspace}
\newcommand{\syst}{\ensuremath{\mathrm{\,(syst)}}\xspace}

\def\gsim{{~\raise.15em\hbox{$>$}\kern-.85em
          \lower.35em\hbox{$\sim$}~}\xspace}
\def\lsim{{~\raise.15em\hbox{$<$}\kern-.85em
          \lower.35em\hbox{$\sim$}~}\xspace}

\def\ptot       {\mbox{$p$}\xspace}
\def\pt         {\mbox{$p_{\rm T}$}\xspace}

\def\evtgen     {\mbox{\textsc{EvtGen}}\xspace}

\def\geant      {\mbox{\textsc{Geant4}}\xspace}

\def\photos     {\mbox{\textsc{Photos}}\xspace}

\def\pythia     {\mbox{\textsc{Pythia}}\xspace}

\def\tell1  {TELL1\xspace}
\def\ukl1   {UKL1\xspace}



%% file: commands.tex
\newcommand{\theresult}{1.84}
\newcommand{\statresult}{0.05}
\newcommand{\systresult}{0.07}
\newcommand{\fdresult}{0.11}
\newcommand{\normresult}{0.12} 

\newcommand{\relresult}{1.84}
\newcommand{\relstatresult}{0.05}
\newcommand{\relsystresult}{0.07}
\newcommand{\relfdresult}{0.11}

%% file: title-LHCb-PAPER.tex

\begin{titlepage}
\pagenumbering{roman}

\vspace*{-1.5cm}
\centerline{\large EUROPEAN ORGANIZATION FOR NUCLEAR RESEARCH (CERN)}
\vspace*{1.5cm}
\hspace*{-0.5cm}
\begin{tabular*}{\linewidth}{lc@{\extracolsep{\fill}}r}
\ifthenelse{\boolean{pdflatex}}
{\vspace*{-2.7cm}\mbox{\!\!\!\includegraphics[width=.14\textwidth]{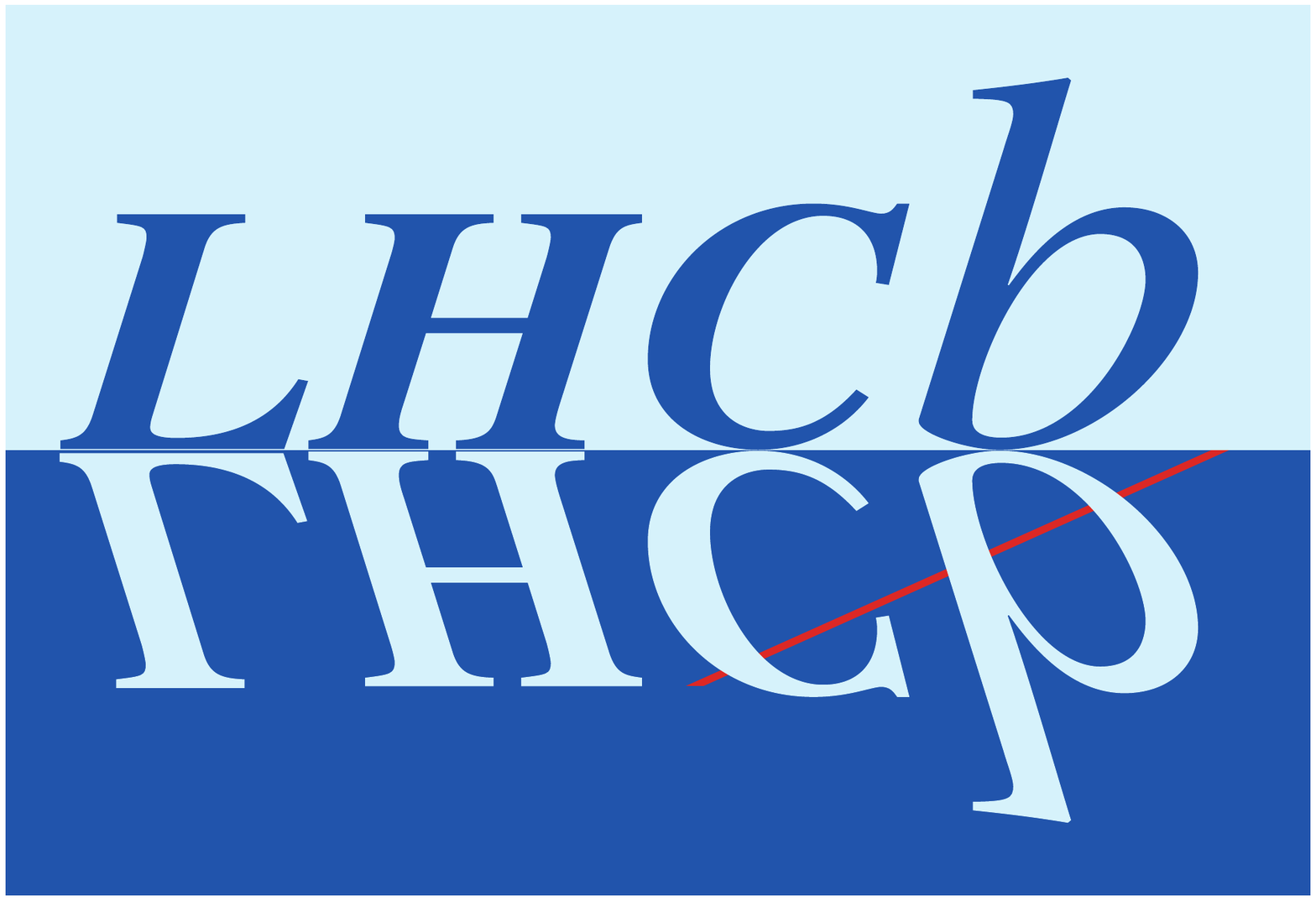}} & &}%
{\vspace*{-1.2cm}\mbox{\!\!\!\includegraphics[width=.12\textwidth]{lhcb-logo.eps}} & &}%
\\
 & & CERN-PH-EP-2015-198 \\  
 & & LHCb-PAPER-2015-028 \\  
 & & 16 October 2015
\end{tabular*}

\vspace*{2.0cm}

{\bf\boldmath\huge
\begin{center}
  Measurement of the $B_s^0 \rightarrow \phi \phi$ branching fraction and search for the decay $B^0 \rightarrow \phi \phi$
\end{center}
}

\vspace*{1.0cm}

\begin{center}
The LHCb collaboration\footnote{Authors are listed at the end of this paper.}
\end{center}

\vspace{\fill}

\begin{abstract}
  \noindent
Using a dataset corresponding to an integrated luminosity of
$3.0\invfb$ collected in $pp$ collisions at centre-of-mass energies of 7 and 8 TeV, the 
$\Bs \rightarrow \phi \phi$ branching fraction is measured to be 
\begin{displaymath}
\mathcal{B}(B^0_s \to \phi \phi) = \left( \theresult \pm \statresult \stat
\pm \systresult \syst \pm \fdresult \, (f_s/f_d) \pm \normresult \, (\text{norm}) \, \right) \times 10^{-5},
\end{displaymath}
where $f_s/f_d$ represents the ratio of the $\Bs$ to $\Bz$ production cross-sections, 
and the $B^0 \to \phi \Kstar(892)^0$ decay mode is used for normalization. 
This is the most precise measurement of this branching fraction to date,
representing a factor five reduction in the statistical uncertainty compared with the previous best measurement.
A search for the decay $\Bz \rightarrow \phi \phi$ is also made. 
No signal is observed, and an upper limit on the branching fraction is set as
\begin{displaymath}
\mathcal{B}(B^0 \to \phi \phi) < 2.8 \times 10^{-8}
\end{displaymath}
at $90 \%$ confidence level. This is a factor of seven improvement compared to the previous best limit.
\end{abstract}

\vspace*{1.0cm}

\begin{center}
  Published as JHEP 10 (2015) 053
\end{center}

\vspace{\fill}

{\footnotesize 
\centerline{\copyright~CERN on behalf of the \lhcb collaboration, license \href{http://creativecommons.org/licenses/by/4.0/}{CC-BY-4.0}.}}
\vspace*{2mm}

\end{titlepage}


\newpage
\setcounter{page}{2}
\mbox{~}

\cleardoublepage

%% file: introduction.tex
\section{Introduction}
\label{sec:Introduction}

In the Standard Model, the flavour-changing neutral current 
decay $\Bs \rightarrow \phi \phi$ proceeds via a $\bar{b}\rightarrow
\bar{s}s\bar{s}$ penguin amplitude. The decay was first observed by the
CDF experiment at the Tevatron \cite{Acosta:2005eu}. Subsequently, it has been studied by the CDF and LHCb
collaborations, who searched for \CP-violating asymmetries in the decay time and angular distributions of 
this mode~\cite{Aaltonen:2011rs,LHCb-PAPER-2013-007,LHCb-PAPER-2012-004,LHCb-PAPER-2014-026}.
These studies provide a probe for possible new physics contributions entering
into the penguin loop and $\Bs - \Bsb$ mixing diagrams~\cite{Raidal:2002ph}. 
Furthermore, as the $\Bs \rightarrow \phi \phi$ mode will be used as normalization for studies of other
charmless $\Bs$ meson decays, it is important to have a precise determination of its branching fraction.
The CDF collaboration measured this relative to the decay $\Bs \rightarrow J/\psi \phi$~\cite{Aaltonen:2011rs}. 
Using the current value of the $\Bs \rightarrow J/\psi \phi$ branching fraction~\cite{PDG2014}, the CDF result 
gives $\mathcal{B}(\Bs \rightarrow \phi \phi) =  (1.91 \pm 0.26 \pm 0.16) \times 10^{-5}$, where the first
uncertainty is from the measured ratio to $\Bs \rightarrow J/\psi \phi$, and the second is due to the knowledge of
the $\Bs \rightarrow J/\psi \phi$ branching fraction.
Various predictions from theories based on QCD factorization exist for the $\Bs \rightarrow \phi \phi$ 
branching fraction~\cite{Beneke:2006hg,Cheng:2009mu,Bartsch:2008ps}.
These suffer from uncertainties related to weak annihilation diagrams. These uncertainties are controlled 
using experimental information from decays such as $\Bz \rightarrow \phi \Kstar(892)^0$. 
Several recent predictions are summarized in Table~\ref{tab:bfpred}. 
The central values are in the range $(1.5 - 2.0) \times 10^{-5}$. 

\begin{table}[b]
\begin{center}
\caption{Predictions for the $\Bs\rightarrow \phi \phi$ branching fraction. 
The first and second uncertainties of Refs.~\cite{Beneke:2006hg,Cheng:2009mu} reflect the knowledge of CKM
parameters and power corrections, respectively.}
{\renewcommand{\arraystretch}{1.2}
\begin{tabular}{l|c|c}
$\mathcal{B}(\Bs \rightarrow \phi \phi)$ ($10^{-5}$) & Approach & Reference \\ \hline
$1.95 \pm 0.10 ^{+1.30}_{-0.80}$ & QCD factorization &\cite{Beneke:2006hg} \\ 
$1.67^{+0.26}_{-0.21}{}^{+1.13}_{-0.88} $ & QCD factorization & \cite{Cheng:2009mu} \\ 
$1.55^{+2.24}_{-1.70}$ & QCD factorization& \cite{Bartsch:2008ps} \\
$1.67^{+0.89}_{-0.71}$ & pQCD & \cite{Zou:2015iwa} \\ 
\end{tabular}}
\label{tab:bfpred}
\end{center}
\end{table}

In this paper the $B_s^0 \rightarrow \phi \phi$ branching fraction (the use of charge-conjugate modes is implied
throughout) is measured using the full LHCb Run 1 dataset, comprising data corresponding to an integrated
luminosity of 1.0\,fb$^{-1}$ collected in $pp$ collisions at a centre-of-mass energy of 7~TeV, and 2.0\,fb$^{-1}$
collected at 8~TeV. 
The decay $B^0 \to \phi \Kstar(892)^0$, which has a similar topology, is used for normalization.
The $\phi$ and $\Kstar(892)^0$ mesons are reconstructed in the $K^+K^-$ and $K^+\pi^-$ final states, respectively.
In addition, a search for the yet unobserved decay $\Bz \rightarrow \phi \phi$ is made.  This decay is suppressed
in the Standard Model by the OZI rule~\cite{Okubo:1963fa, *Zweig:1964, *Iizuka:1966fk},  with an expected branching 
fraction in the range $(0.1 - 3.0) \times 10^{-8}$~\cite{Lu:2005be, BarShalom:2002sv, Beneke:2006hg, Bartsch:2008ps}.  
However, the branching fraction can be enhanced, up to the $10^{-7}$ level, in models such as
supersymmetry with R-parity violation~\cite{BarShalom:2002sv}. 
The current best limit for this mode is from the BaBar collaboration~\cite{Aubert:2008fq}, 
$\mathcal{B}( \Bz \rightarrow \phi \phi) < 2.0 \times 10^{-7}$ at $90 \, \%$ confidence level.

%% file: detector.tex
\section{Detector and software}
\label{sec:Detector}

The \lhcb detector~\cite{Alves:2008zz,LHCb-DP-2014-002} is a single-arm forward
spectrometer covering the \mbox{pseudorapidity} range $2<\eta <5$,
designed for the study of particles containing \bquark or \cquark
quarks. The detector includes a high-precision tracking system
consisting of a silicon-strip vertex detector surrounding the $pp$
interaction region~\cite{LHCb-DP-2014-001}, a large-area silicon-strip detector located
upstream of a dipole magnet with a bending power of about
$4{\rm\,Tm}$, and three stations of silicon-strip detectors and straw
drift tubes~\cite{LHCb-DP-2013-003}  placed downstream of the magnet.
The tracking system provides a measurement of momentum, \ptot,  with
a relative uncertainty that varies from $0.5 \, \%$ at low momentum to
$1.0 \, \%$ at 200\gevc. The minimum distance of a track to a primary $pp$ interaction vertex~(PV), the
impact parameter, is measured with a resolution of $(15+29/\pt)\mum$,
 where \pt is the component of the momentum transverse to the beam, in\,\gevc.

Different types of charged hadrons are distinguished using information
from two ring-imaging Cherenkov detectors~\cite{LHCb-DP-2012-003}. Photon, electron and
hadron candidates are identified by a calorimeter system consisting of
scintillating-pad and preshower detectors, an electromagnetic
calorimeter and a hadronic calorimeter. Muons are identified by a
system composed of alternating layers of iron and multiwire
proportional chambers~\cite{LHCb-DP-2012-002}. 

The trigger~\cite{LHCb-DP-2012-004}  consists of a
hardware stage, based on information from the calorimeter and muon
systems, followed by a software stage, which applies a full event
reconstruction. The software trigger applied in this analysis requires a two-, three- or four-track
secondary vertex with a significant displacement from any PV.
At least one charged particle must have a transverse momentum $\pt > 1.7\gevc$ and be
inconsistent with originating from a PV.
A multivariate algorithm~\cite{BBDT} is used for
the identification of secondary vertices consistent with the decay
of a \bquark hadron.

In the simulation, $pp$ collisions are generated using
\pythia~\cite{Sjostrand:2006za,*Sjostrand:2007gs}  with a specific \lhcb
configuration~\cite{LHCb-PROC-2010-056}.  Decays of hadronic particles
are described by \evtgen~\cite{Lange:2001uf}, in which final-state
radiation is generated using \photos~\cite{Golonka:2005pn}. The
interaction of the generated particles with the detector, and its
response, are implemented using the \geant
toolkit~\cite{Allison:2006ve, *Agostinelli:2002hh} as described in
Ref.~\cite{LHCb-PROC-2011-006}.

%% file: selection.tex
\section{Signal selection}
\label{sec:selection}

The selection of candidates takes place in two stages. First, a
selection using loose criteria is performed that reduces
background whilst retaining high signal efficiency. Following this, a multivariate method
is used to further improve the signal significance.

The selection starts from charged particle tracks that traverse the entire spectrometer. 
Selected particles are required to have $\pt > 500 \, \mevc$. 
Fake tracks created by the reconstruction due to random combinations of hits in the detector are suppressed 
using a requirement on a neural network trained to discriminate between these and genuine tracks
associated to particles.
Combinatorial background from hadrons originating at the primary vertex is suppressed by requiring that 
all tracks are significantly displaced from any primary vertex. 
Kaon and pion candidates are selected using the information provided by the ring-imaging Cherenkov detectors. 
This is combined with kinematic information using a neural network to provide an effective probability 
that a particle is a kaon ($\mathcal{P}^K$) or pion ($\mathcal{P}^{\pi}$). 
To select kaon candidates it is required that $\mathcal{P}^K (1 - \mathcal{P}^{\pi}) > 0.025$. 
The pion candidate in the $\Bz \to \phi \Kstar(892)^0$ decay mode is required
to have $\mathcal{P}^{\pi} > 0.2$ and $\mathcal{P}^K < 0.2$. 

The selected charged particles are combined to form $\phi$ and $K^*$ meson candidates. 
The invariant mass of the $K^+K^-$ ($K^+\pi^-)$ pair is required to be within $15 \mevcc$ ($150 \mevcc$) of 
the known mass of the $\phi$ ($K^*(892)^0$) meson~\cite{PDG2014}.
In addition, the $\pt$ of the $\phi$ and $K^*$ mesons must be greater than $1 \gevc$. 

Candidates for the decay $\Bs \rightarrow \phi \phi$ are formed by combining pairs of $\phi$ mesons. 
A fit is made requiring all four final-state particles to originate from a common vertex, and
the direction vector between the primary and secondary vertices is required to be consistent with the
direction of the momentum vector of the $\Bs$ meson candidate. 
Further requirements are then applied to remove background from specific $\bquark$-hadron decays that peak close to
the $\Bs$ mass. 
To reject background from $\Bz \to \phi \Kstar(892)^0$ decays, the kaon with the lowest value of $\mathcal{P}^{K}$ 
is considered to be a pion, and the $K^+ \pi^-$ and $K^+ K^-K^+ \pi^-$ invariant masses are calculated. 
Candidates with $m(K^+ \pi^-)$ within $ 50 \mevcc$ of the known $K^*(892)^0$ mass and $m(K^+ K^-K^+ \pi^-)$ within 
$30 \mevcc$ of the $\Bz$ mass~\cite{PDG2014} are rejected. 
Similarly, to remove decays via open charm mesons, the $K^+K^-\pi^+$ mass is calculated. 
If $m(K^+K^-\pi^+)$ is within $22.5 \mevcc$ of the $D^+$ or $D^+_s$ mass~\cite{PDG2014}, the candidate is rejected.
These vetoes are found to retain 91\% of simulated $\Bs \rightarrow \phi \phi$ decays.

Candidates for the decay $\Bz \to \phi \Kstar(892)^0$ are formed from combinations of $\phi$ and $\Kstar$ mesons. 
Identical vertex and pointing requirements as for the $\Bs \rightarrow \phi \phi$ decay mode are applied.  
To reject background from $\Bs \to \phi \phi$, the mass of the $K^+ \pi^-$ pair is calculated assuming that both 
hadrons are kaons. Candidates with $m(K^+ K^-)$ within $15\mevcc$ of the $\phi$ mass and $m(K^+ K^- K^+ K^-)$ 
within $30 \mevcc$ of the $\Bs$ mass~\cite{PDG2014} are rejected. 
Background from open charm decays is suppressed in a manner similar to that used for 
the $\Bs \rightarrow \phi \phi$ candidates.
These vetoes are found to retain 97\% of simulated $\Bz \to \phi \Kstar(892)^0$ decays.

The combinatorial background is further suppressed using a Boosted Decision Tree method (BDT)~\cite{Breiman,AdaBoost}.  
The BDT is trained to identify four-body hadronic $b$-hadron decays with high efficiency using independent data samples of such decays.
It uses information on the displacement of the $b$-hadron candidate from the primary vertex, kinematic information 
and track isolation criteria. 
Although the same BDT is used for the $\Bs \rightarrow \phi \phi$ branching fraction measurement and the
search for $\Bz \rightarrow \phi \phi$, the method used to optimize the cut on the BDT output is different.
For the branching fraction measurement, the cut optimization is based on the normalisation mode $\Bz \to \phi \Kstar(892)^0$.
The figure of merit used is 
\begin{equation}
\frac{S_0 \times \varepsilon_{S}}{\sqrt{S_0 \times \varepsilon_{S}  + N_{\text{bg}}}}, \nonumber
\label{eq:Bs2phiphiFOM}
\end{equation}
where $S_0$ is the signal yield of $\Bz \to \phi \Kstar(892)^0$ candidates in data before any BDT cut is applied, $\varepsilon_{S}$ is 
the efficiency of the BDT cut on simulated $\Bz \to \phi \Kstar(892)^0$ decays, and $N_{\text{bg}}$ is the number of background candidates surviving 
the BDT cut in a suitable upper sideband of the $\phi \Kstar(892)^0$ candidate mass distribution, scaled to the width of the $\Bz$ signal window.
Maximizing this figure of merit results in a rather loose BDT requirement that retains $98 \%$ of signal events 
while rejecting more than 90\% of the background.

For the $\Bz \rightarrow \phi \phi$ search, the figure of merit used is
\begin{equation}
\frac{\varepsilon^{\prime}_{S}}{a/2+\sqrt{N^{\prime}_{\text{bg}}}}, \nonumber
\label{eq:Bd2phiphiFOM}
\end{equation}
with $a$ set to 3, corresponding to the signal significance required to claim evidence for a new decay mode~\cite{Punzi:2003bu}.
Here $\varepsilon^{\prime}_{S}$ is the efficiency of the BDT cut on simulated $\Bs \rightarrow \phi \phi$ decays, and 
$N^{\prime}_{\text{bg}}$ is the number of background candidates surviving the BDT cut in an upper sideband of the $\phi \phi$ 
candidate mass distribution, scaled to the width of the $\Bs$ signal window.
Maximizing this figure of merit results in a tighter BDT requirement that retains $87 \%$ of signal events.

%% file: massfit.tex
\section{Fits to mass spectra}
\label{sec:mass}

\begin{figure}[h]
\centering
\includegraphics[width=0.7\textwidth]{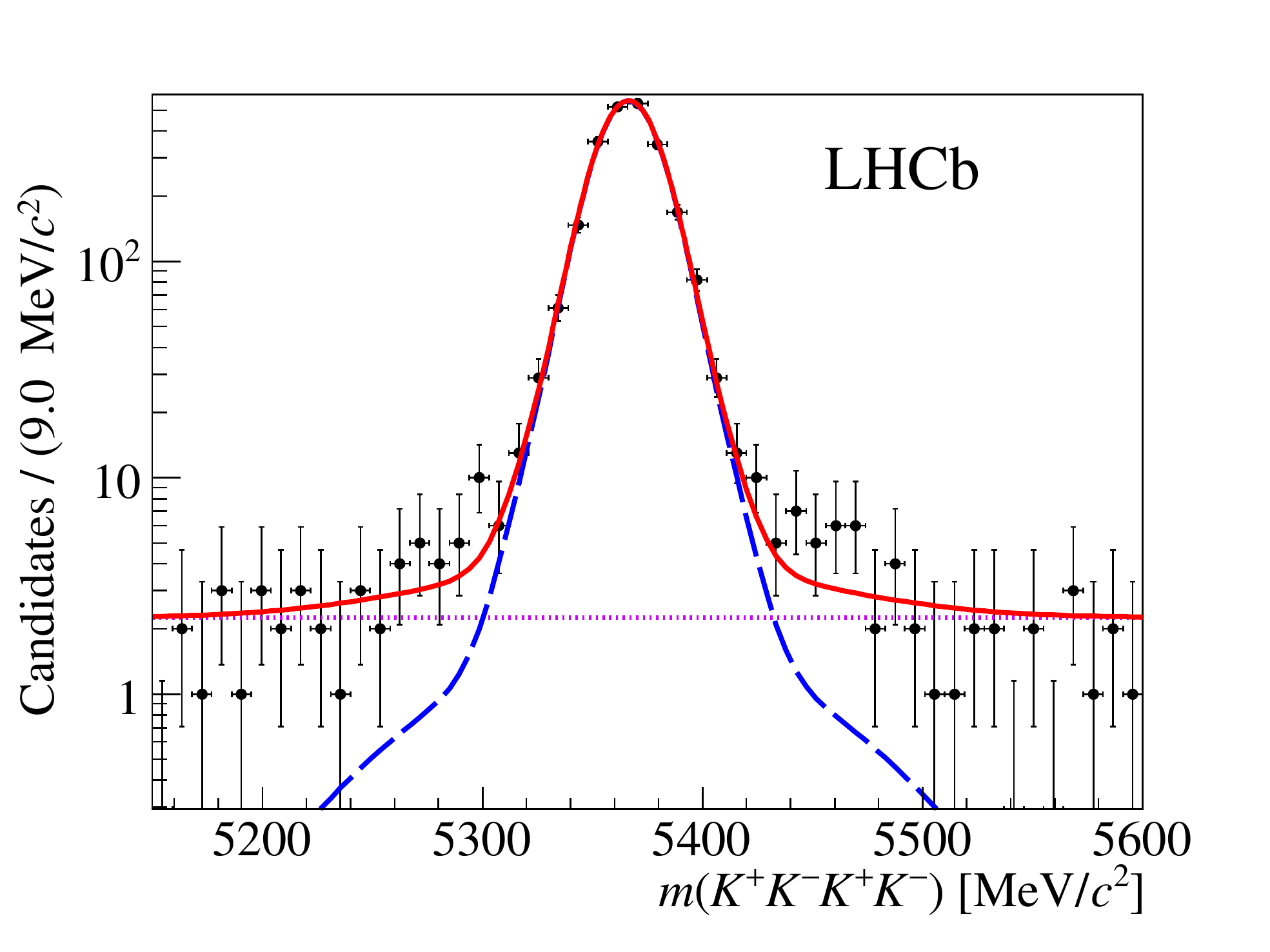}
\caption{The $K^+K^-K^+K^-$ invariant mass distribution.
  The total fitted function as described in the text is shown by the (red) solid line,
  the $\Bs\to\phi\phi$ component by the (blue) long-dashed line,
  and the combinatorial background as the (purple) dotted line.
  }
\label{fig:phiphimass}
\end{figure}

The yields for the signal and normalization channels are determined from
fits to the invariant mass distributions of the selected candidates. In the simulation, the $\Bs \rightarrow \phi
\phi$ invariant mass distribution is well modelled 
by a probability density function (PDF) consisting of the sum of three Gaussian distributions with a common mean. 
In the fit to the data, the relative fractions of the Gaussian components are fixed to the values obtained from the simulation, 
whilst the widths are allowed to vary by an overall resolution scale factor. The yield and common mean are also left free.
After applying all selection requirements, the only remaining background is combinatorial, which is modelled by a constant. 
No component for $\Bz \rightarrow \phi \phi$ decays is included in this fit.
Figure~\ref{fig:phiphimass} shows the resulting fit to data, which gives a signal yield of $2309\pm49$ candidates.

The $\Bz \to \phi \Kstar(892)^0$ invariant mass distribution is modelled by a PDF consisting of the sum of a 
Crystal Ball function~\cite{Skwarnicki:1986xj} and two Gaussian functions.
As for the signal mode, the relative fractions of the components and the tail parameters are fixed in the fit to the data,
whilst the widths are allowed to vary by an overall resolution scale factor. The yield and mean are also left free.
A component is also included to account for the small contribution from the decay 
$\Bs \to \phi \Kstarb(892)^0$~\cite{LHCb-PAPER-2013-012}. The shape parameters for this component are shared with the 
$\Bz$ component, while the relative position is fixed to the known mass difference between the \Bz and \Bs mesons~\cite{PDG2014}. 
Combinatorial background is modelled by an exponential function. 

\begin{figure}[h]
\centering
\includegraphics[width=0.7\textwidth]{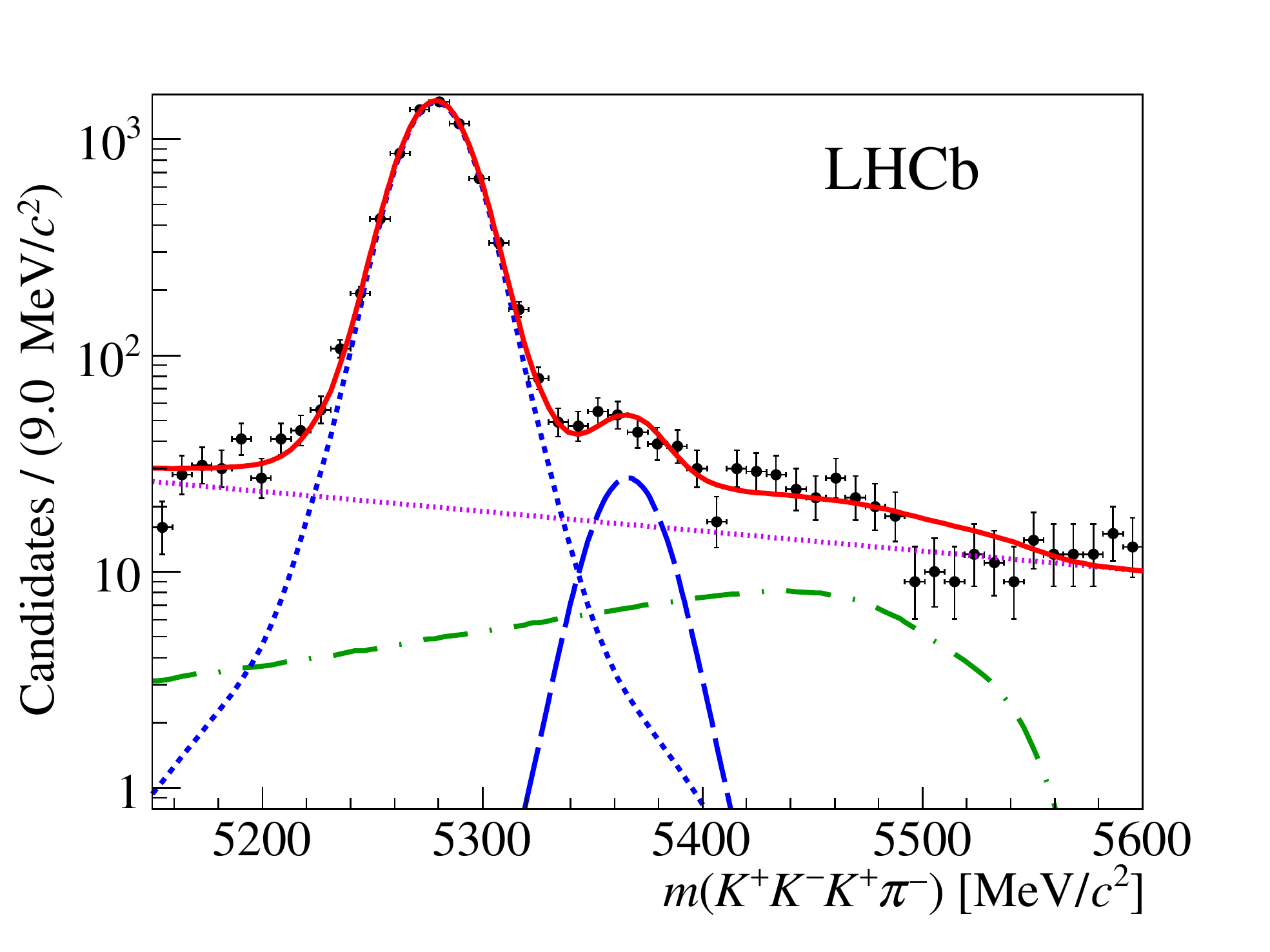}
\caption{
  The $K^+K^-K^+\pi^-$ invariant mass distribution.
  The total fitted function is shown by the (red) solid line,
  the $\Bd\to\phi\Kstar$ component by the (blue) short-dashed line,
  the $\Bs\to\phi\Kstarb(892)^0$ component by the (blue) long-dashed line,
  the $\Lb\to\phi\proton\pi^-$ and $\Lb\to\phi\proton K^-$ contribution by the (green) dashed-dotted line,
  and the combinatoral background by the (purple) dotted line.
  }
\label{fig:phikstmass}
\end{figure}

Potential peaking backgrounds, from $\Lb\to\phi\proton\pi^-$ (with the proton misidentified as a kaon) or 
$\Lb\to\phi\proton K^-$ (with the proton misidentified as a pion), are modelled using a single histogram PDF generated
from simulated events. The relative yield of each decay mode is weighted according to the expectation from the simulation.
The yield of this component is left to float in the fit. 
Backgrounds from $\Bs \to \phi \phi$ and open charm decay modes are negligible after the vetoes described in 
Section~\ref{sec:selection} have been applied.
Figure~\ref{fig:phikstmass} shows the result of the fit of this model to the $\Bz \to\phi \Kstar(892)^0$ dataset 
after all selection criteria are applied.
The yield of $\Bz$ candidates determined by the fit is $6680\pm86$.

%% file: systematics.tex
\section{Branching fraction for $\Bs \to \phi \phi$}
\label{sec:syst}

The branching fraction of $\Bs \to \phi \phi$ relative to that of the $\Bd \to \phi \Kstar(892)^0$ decay mode is determined using
\begin{equation}
  \label{eq:phiphiBR}
  \frac{\mathcal{B}(\Bs \to \phi \phi)}{\mathcal{B}(\Bd \to \phi \Kstar(892)^0)} = 
  \frac{N_{\phi \phi}}{N_{\phi \Kstar(892)^0}}
  \frac{\varepsilon^\text{sel}_{\phi \Kstar(892)^0}}{\varepsilon^\text{sel}_{\phi \phi}}
  \frac{\mathcal{B}(\Kstar(892)^0 \to K^+ \pi^-)}{\mathcal{B}(\phi \to K^+ K^-)}
  \cdot
  \frac{1}{f_s/f_d}, \nonumber
\end{equation}
where the terms $\mathcal{B}$ are the branching fractions of the stated decay modes, $N$ are the signal yields, $\varepsilon^\text{sel}$ are the selection efficiencies, and the fragmentation fraction ratio, $f_s/f_d$, is the ratio of the $\Bs$ to $\Bz$ production cross-sections.
The selection efficiencies are determined from simulation, apart from those related to the particle identification, which are determined in data using large calibration samples of charged kaons and pions from $\Dstar^+ \to \D^0(\to K^-\pi^+) \pi^+$ decays~\cite{LHCb-DP-2012-003}.
The ratio of efficiencies is found to be
$\varepsilon^\text{sel}_{\phi\Kstar(892)^0}/\varepsilon^\text{sel}_{\phi \phi} = 0.795 \pm 0.007$, where the uncertainty is purely statistical.
The value of $f_s/f_d$ is taken from previous LHCb analyses as $0.259 \pm 0.015$~\cite{LHCb-PAPER-2011-018,LHCb-PAPER-2012-037, LHCb-CONF-2013-011}.

The signal yields are determined using the mass fits described in Section~\ref{sec:mass}.
These values are corrected for the fraction of candidates where one of the hadron pairs, $K^+K^-$ or $K^+\pi^-$, is produced in a non-resonant S-wave configuration, rather than as a $\phi$ or $\Kstar(892)^0$.
The S-wave fractions are taken from previous LHCb angular analyses of the $\Bs\to \phi \phi$ and $\Bd \to \phi\Kstar(892)^0$ decay modes.
For the $\Bs\rightarrow \phi \phi$ decay mode, we use the measured value of $2.1\pm1.6\, \%$~\cite{LHCb-PAPER-2014-026} as the S-wave fraction within the $K^+K^-$ invariant mass range used for this analysis.
Similarly, for the $\Bd \to \phi\Kstar(892)^0$ decay mode, we use a measured value of $26.5 \pm 1.8 \, \%$~\cite{LHCb-PAPER-2014-005} for the S-wave fraction.
The uncertainties on these fractions lead to a $3.1 \, \%$ relative uncertainty on the ratio of branching fractions.
This procedure assumes that the efficiencies for the P- and S-wave components are the same.
In the simulation, a $1.1 \, \% $ difference is observed between these efficiencies, and this is assigned as an additional uncertainty.

Various other uncertainties arise on the measurement of the ratio of branching fractions.
The limited size of the available simulation samples leads to a relative uncertainty of $0.8 \, \%$.
The influence of the assumed mass model is probed by performing the fit with different models for the signal and background components.
This includes quantifying the effect of removing the peaking background component in the $\Bd \to \phi \Kstar(892)^0$ fit.
The largest variation in the ratio of branching fractions seen in these studies is $0.6 \, \%$, which is assigned as a relative systematic uncertainty. 

The track reconstruction efficiency agrees between data and simulation at the level of $2.0  \, \%$~\cite{LHCb-DP-2013-002}.
This uncertainty largely cancels in the ratio of branching fractions.
A residual relative uncertainty of $0.5  \, \%$ remains due to the fact that the pion in the $\Bd \to \phi \Kstar(892)^0$ decay mode is relatively soft.
An additional relative uncertainty of  $0.3 \, \%$ is assigned to account for the difference in the hadronic interaction probabilities for kaons and pions between data and simulation.
A further uncertainty arises from the modelling of the hardware trigger in the simulation.
This is estimated using a data-driven technique and leads to a relative systematic uncertainty of $1.1  \, \%$ on the ratio of branching fractions.
Variations in the procedure used to determine the relative particle identification efficiency lead to a relative uncertainty of $0.3  \, \%$.
Possible systematic effects on the efficiency for $\Bs \to \phi \phi$ due to the finite width difference in the $\Bs$ system~\cite{DeBruyn:2012wj} have 
been checked, and found to be negligible.
The value of $\mathcal{B}(\phi\to K^+ K^-)$ is taken from Ref.~\cite{PDG2014} and contributes a relative uncertainty of $1.0 \, \%$.
The value of $\mathcal{B}(\Kstar(892)^0\to K^+ \pi^-)$ is taken to be $2/3$ exactly.
The systematic uncertainties are summarized in Table~\ref{tab:BRsyst}.
Summing these in quadrature gives a relative uncertainty of $3.8 \, \%$ on the ratio of branching fractions.
The knowledge of the fragmentation fraction ratio, $f_s/f_d$, gives a relative uncertainty of $5.8  \, \%$, which is quoted separately. 
\begin{table}[t]
  \caption{Summary of the systematic uncertainties on the measurement of the ratio of branching fractions $\mathcal{B}(\Bs \to \phi \phi)/\mathcal{B}(\Bd \to \phi  K^*)$.}
  \begin{center}
    \begin{tabular}{c|c}
      Source of systematic uncertainty          & Relative uncertainty (\%)\\  \hline
      S-wave fraction                           & 3.1 \\          
      Relative efficiency between P and S-wave  & 1.1 \\
      Simulation sample size                    & 0.8 \\
      Fit model                                 & 0.6 \\
      Tracking efficiency                       & 0.5 \\
      Hadronic interactions                     & 0.3 \\
      Hardware trigger                          & 1.1 \\
      Particle identification efficiency        & 0.3 \\
      $\mathcal{B}(\phi \rightarrow K^+ K^-)$   & 1.0 \\
      \hline
      Quadratic sum of the above                & 3.8 \\          
      \hline
      Fragmentation fraction ratio ($f_s/f_d$)  & 5.8
    \end{tabular}
  \end{center}
  \label{tab:BRsyst}
\end{table}

The ratio of branching fractions is found to be
\begin{displaymath}
  \frac{\mathcal{B}(\Bs \to \phi \phi)}{\mathcal{B}(\Bd \to \phi  K^*)} =
  \relresult \pm \relstatresult \stat\pm \relsystresult \syst\pm \relfdresult \, (f_s/f_d).
\end{displaymath}
This is converted into an absolute branching fraction using $\mathcal{B}(\Bd \to \phi \Kstar(892)^0) = (1.00 \pm 0.04\pm0.05)\times10^{-5}$, which is obtained by averaging the results in Refs.~\cite{Aubert:2008zza} and~\cite{Prim:2013nmy} assuming that the uncertainties due to the fragmentation fractions and S-waves are fully correlated between the two measurements.
The resulting value for the absolute branching fraction is
\begin{displaymath}
  \mathcal{B}(\Bs \to \phi \phi) =
  \left(
  \theresult \pm \statresult \stat \pm \systresult \syst \pm
  \fdresult \, (f_s/f_d) \pm \normresult \, (\text{norm})
  \right)
  \times 10^{-5}.
\end{displaymath}

%% file: bdlimit.tex
\section{Search for the decay $\Bd \rightarrow \phi \phi$}
\label{sec:phi}

\begin{figure}[b]
\centering
\includegraphics[width=0.7\textwidth]{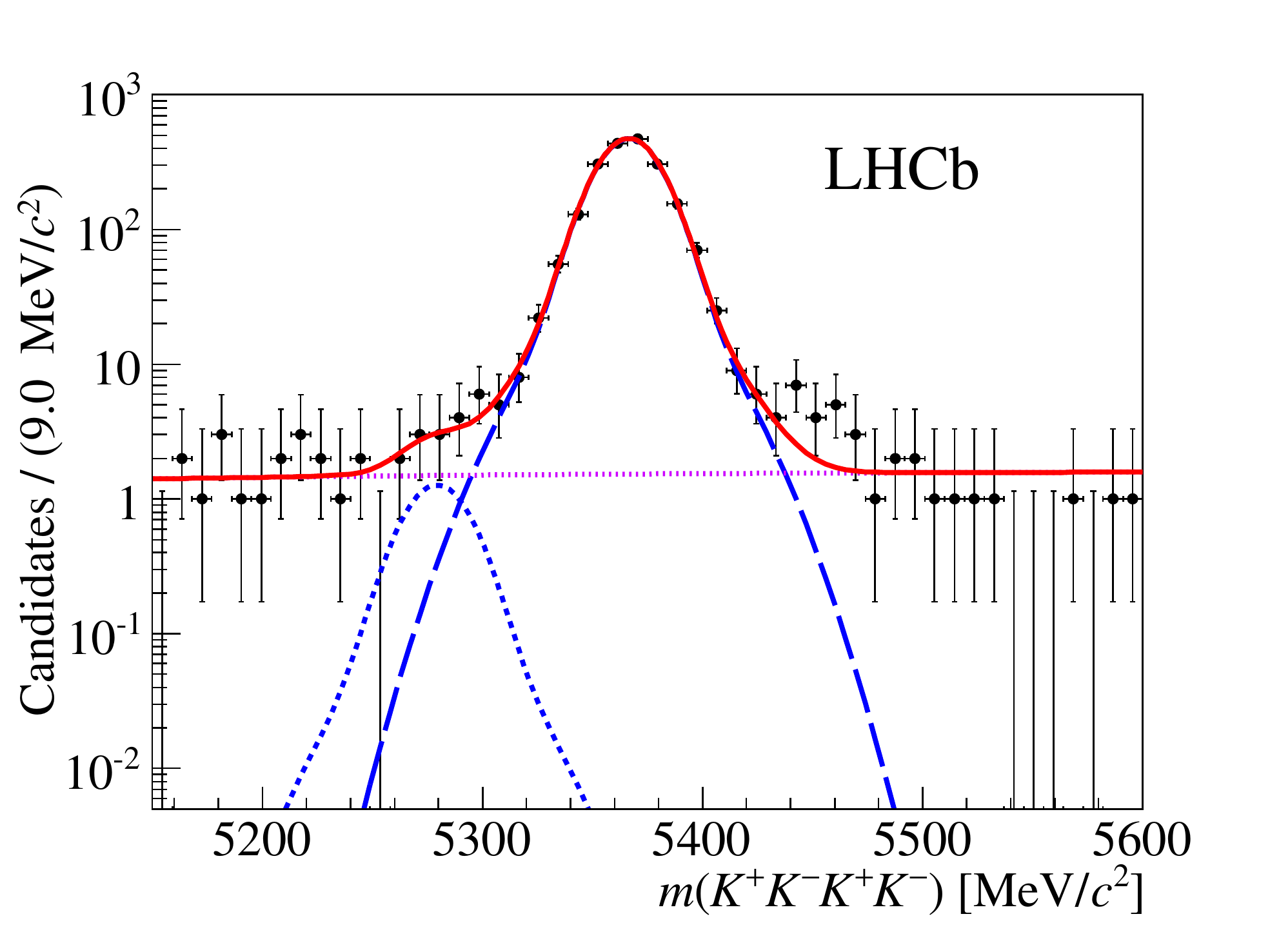}
\caption{ The $K^+K^-K^+K^-$ invariant mass with the tight BDT selection applied.
  A fit to the total PDF as described in the text is shown as a (red) solid line,
  $\Bs\to\phi\phi$ as a (blue) long-dashed line,
  $\Bd\to\phi\phi$ as a (blue) short-dashed line,
  and the combinatorial background as a (purple) dotted line.}
\label{fig:phiphi_wBdmass}
\end{figure}

To search for the $\Bd \rightarrow \phi \phi$ decay mode, the tight BDT selection described in Section~\ref{sec:selection} is used.
To fit for a putative $\Bd \rightarrow \phi \phi$ signal, the same signal model as for the $\Bs$ signal is used.
The mean value of the signal mass is shifted relative to the $\Bs$ mode by the known \Bs--\Bd mass splitting, and the resolution parameters are kept common between the two modes.
The resulting fit is shown in Figure~\ref{fig:phiphi_wBdmass}.
The data are consistent with having no $\Bd\to\phi\phi$ contribution.
The fitted $\Bd$ signal has a yield of $5\pm6$ events, and the statistical significance is less than 2 standard deviations, hence an upper limit is placed on the branching fraction of the decay. 

\begin{figure}[t]
\centering
\includegraphics[width=0.7\textwidth]{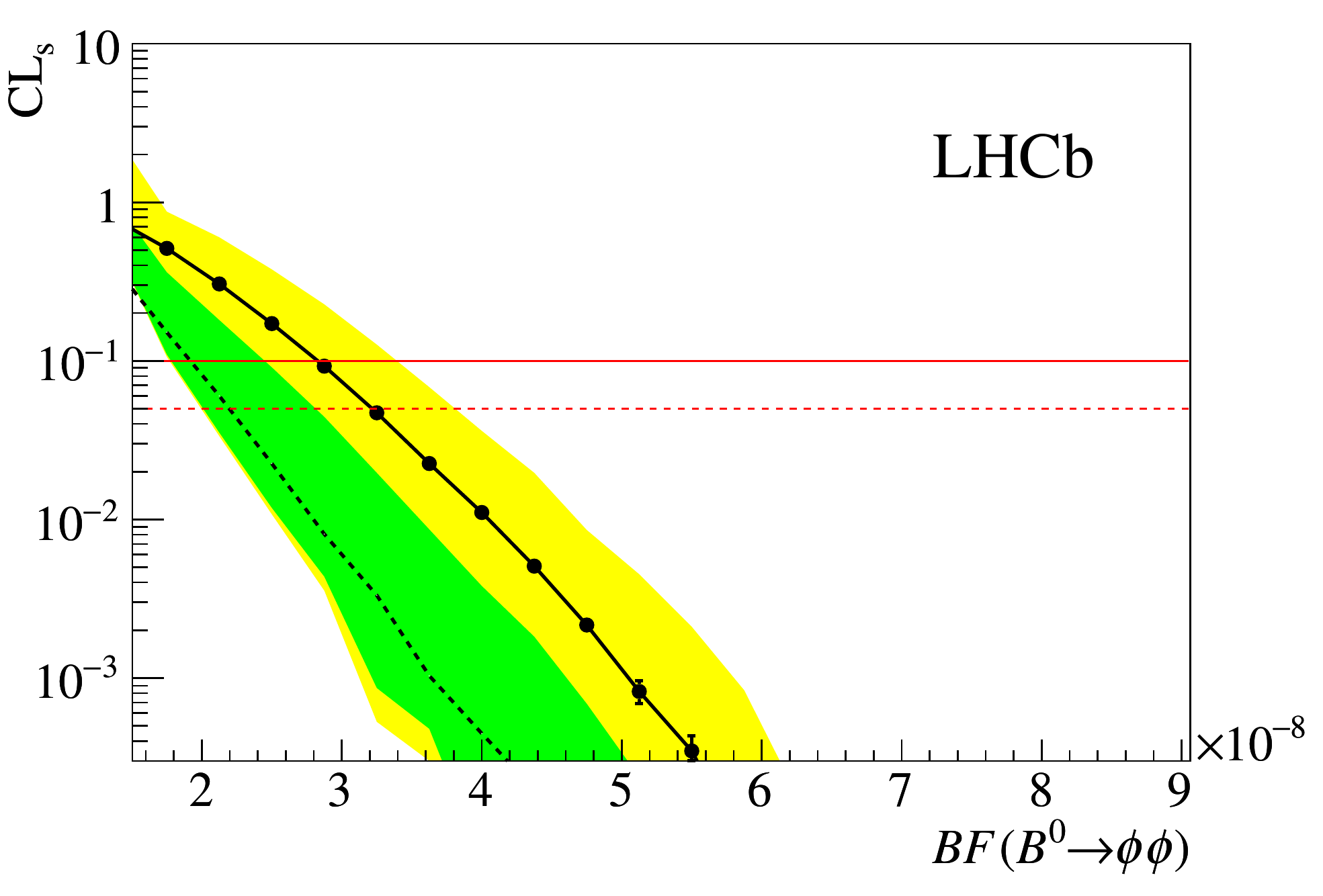}
\caption{
Results of the  $\text{CL}_\text{s}$ scan as a function of the $\Bd\to\phi\phi$ branching fraction ($BF$).
The observed $\text{CL}_\text{s}$ distribution is given by the (black) points and solid line,
while the expected distribution is given by the (black) dashed line.
The dark (green) and light (yellow) bands mark the $1\sigma$ and $2\sigma$ confidence regions on the expected $\text{CL}_\text{s}$.
The upper limits at 90$\,\%$ and 95$\,\%$ confidence level are where the observed $\text{CL}_\text{s}$ line intercepts the (red) solid and dashed horizontal lines, respectively.
}
\label{fig:clsfinal}
\end{figure}

To determine this limit, a modified frequentist approach, the $\text{CL}_\text{s}$ method, is used~\cite{Read:2002hq}.
The method provides $\text{CL}_\text{s+b}$, a measure of the compatibility of the observed distribution with the signal plus background hypothesis, $\text{CL}_\text{b}$, a measure of the compatibility with the background only hypothesis, and $\text{CL}_\text{s} = \text{CL}_\text{s+b}/\text{CL}_\text{b}$.
The expected and observed $\text{CL}_\text{s}$ values as a function of the branching fraction are shown in Figure~\ref{fig:clsfinal}.
This gives, at $90 \%$ confidence level, an upper limit of $\mathcal{B}(\Bd \to \phi \phi) < 2.8 \times 10^{-8}$. At $95 \%$ confidence level, the upper limit is found to be $\mathcal{B}(\Bd \to \phi \phi) < 3.4 \times 10^{-8}$.

%% file: summary.tex
\section{Summary}
\label{sec:summary}

The ratio of branching fractions $\mathcal{B}(\Bs \to \phi \phi)/\mathcal{B}(\Bd \to \phi  K^*)$ is determined to be
\begin{displaymath}
  \frac{\mathcal{B}(\Bs \to \phi \phi)}{\mathcal{B}(\Bd \to \phi  K^*)} =
  \relresult \pm \relstatresult \stat\pm \relsystresult \syst\pm \relfdresult \, (f_s/f_d),
\end{displaymath}
where the first uncertainty is statistical, the second is systematic, and the third is due to the ratio of fragmentation fractions.
The absolute branching fraction for $\Bs \to \phi \phi$ is determined to be 
\begin{displaymath}
  \mathcal{B}(\Bs \to \phi \phi) =
  \left(
  \theresult \pm \statresult \stat \pm \systresult \syst \pm
  \fdresult \, (f_s/f_d) \pm \normresult \, (\text{norm})
  \right)
  \times 10^{-5}.
\end{displaymath}
This is in agreement with, but more precise than, the measurement made by the CDF collaboration, ${\mathcal{B}(\Bs \rightarrow \phi \phi) =  \left(1.91 \pm  0.26 \pm 0.16\right) \times 10^{-5}}$. It is also in agreement with theory predictions \cite{Beneke:2006hg,Cheng:2009mu,Bartsch:2008ps,Zou:2015iwa}. 

A search for the decay $\Bd \rightarrow \phi \phi$ is also made.
No significant signal is seen, and an upper limit of
\begin{displaymath}
\mathcal{B}( \Bd \rightarrow \phi \phi) < 2.8 \times 10^{-8}
\end{displaymath}
is set at $90 \%$ confidence level.
This is more stringent than the previous limit of ${\mathcal{B}(\Bd\rightarrow\phi\phi)<2.0\times 10^{-7}}$, set by BaBar \cite{Aubert:2008fq}, and provides a strong constraint on possible contributions to this mode from physics beyond the Standard Model \cite{BarShalom:2002sv}.

%% file: acknowledgements.tex
\section*{Acknowledgements}
 
\noindent We express our gratitude to our colleagues in the CERN
accelerator departments for the excellent performance of the LHC. We
thank the technical and administrative staff at the LHCb
institutes. We acknowledge support from CERN and from the national
agencies: CAPES, CNPq, FAPERJ and FINEP (Brazil); NSFC (China);
CNRS/IN2P3 (France); BMBF, DFG, HGF and MPG (Germany); INFN (Italy); 
FOM and NWO (The Netherlands); MNiSW and NCN (Poland); MEN/IFA (Romania); 
MinES and FANO (Russia); MinECo (Spain); SNSF and SER (Switzerland); 
NASU (Ukraine); STFC (United Kingdom); NSF (USA).
The Tier1 computing centres are supported by IN2P3 (France), KIT and BMBF 
(Germany), INFN (Italy), NWO and SURF (The Netherlands), PIC (Spain), GridPP 
(United Kingdom).
We are indebted to the communities behind the multiple open 
source software packages on which we depend. We are also thankful for the 
computing resources and the access to software R\&D tools provided by Yandex LLC (Russia).
Individual groups or members have received support from 
EPLANET, Marie Sk\l{}odowska-Curie Actions and ERC (European Union), 
Conseil g\'{e}n\'{e}ral de Haute-Savoie, Labex ENIGMASS and OCEVU, 
R\'{e}gion Auvergne (France), RFBR (Russia), XuntaGal and GENCAT (Spain), Royal Society and Royal
Commission for the Exhibition of 1851 (United Kingdom).

%% file: LHCb_authorlist.tex
\centerline{\large\bf LHCb collaboration}
\begin{flushleft}
\small
R.~Aaij$^{38}$, 
B.~Adeva$^{37}$, 
M.~Adinolfi$^{46}$, 
A.~Affolder$^{52}$, 
Z.~Ajaltouni$^{5}$, 
S.~Akar$^{6}$, 
J.~Albrecht$^{9}$, 
F.~Alessio$^{38}$, 
M.~Alexander$^{51}$, 
S.~Ali$^{41}$
% ...

{\footnotesize \it
% ...
$ ^{5}$Clermont Universit\'{e}, Universit\'{e} Blaise Pascal, CNRS/IN2P3, LPC, Clermont-Ferrand, France\\
$ ^{6}$CPPM, Aix-Marseille Universit\'{e}, CNRS/IN2P3, Marseille, France\\
% ...
$ ^{9}$Fakult\"{a}t Physik, Technische Universit\"{a}t Dortmund, Dortmund, Germany\\
% ...
$ ^{37}$Universidad de Santiago de Compostela, Santiago de Compostela, Spain\\
$ ^{38}$European Organization for Nuclear Research (CERN), Geneva, Switzerland\\
% ...
$ ^{41}$Nikhef National Institute for Subatomic Physics, Amsterdam, The Netherlands\\
% ...
$ ^{46}$H.H. Wills Physics Laboratory, University of Bristol, Bristol, United Kingdom\\
% ...
$ ^{51}$School of Physics and Astronomy, University of Glasgow, Glasgow, United Kingdom\\
$ ^{52}$Oliver Lodge Laboratory, University of Liverpool, Liverpool, United Kingdom\\
% ...
}
\end{flushleft}